\documentclass[%
 reprint,
 superscriptaddress,
 amsmath,amssymb,
 aps,
]{revtex4-1}

\usepackage{graphicx}
\usepackage{dcolumn}
\usepackage{bm}

\begin{document}


\title{Fluctuation-induced Distributed Resonances in Oscillatory Networks}

\author{Xiaozhu Zhang}
 \email{xiaozhu.zhang@tu-dresden.de}
\affiliation{%
Chair for Network Dynamics, Institute for Theoretical Physics and Center for Advancing Electronics Dresden (cfaed), Technical University of Dresden, 01062 Dresden, Germany.
}%

\author{Sarah Hallerberg}%
\affiliation{%
Faculty for Engineering and Computer Science, Hamburg University of Applied Science, 20099 Hamburg, Germany.
}%

\author{Moritz Matthiae}
\affiliation{
Institute for Energy and Climate Research - Systems Analysis and Technology Evaluation (IEK-STE), Forschungszentrum J\"ulich GmbH, 52425 J\"ulich, Germany.
}%

\author{Dirk Witthaut}
\affiliation{
Institute for Energy and Climate Research - Systems Analysis and Technology Evaluation (IEK-STE), Forschungszentrum J\"ulich GmbH, 52425 J\"ulich, Germany.
}%
\affiliation{
 Institute for Theoretical Physics, University of Cologne, 50923 K\"oln, Germany.
}%
\author{Marc Timme}
\email{marc.timme@tu-dresden.de}
\affiliation{%
Chair for Network Dynamics, Institute for Theoretical Physics and Center for Advancing Electronics Dresden (cfaed), Technical University of Dresden, 01062 Dresden, Germany.
}%
\affiliation{%
 Department of Physics, Technical University of Darmstadt, 64289 Darmstadt, Germany.
}%

\date{\today}

\begin{abstract}
Across physics, biology and engineering, the collective dynamics of oscillatory networks often evolve into self-organized operating states.
How such networks respond to external fluctuating signals fundamentally underlies their function, yet is not well understood. 
Here we present a theory of dynamic network response patterns, and reveal how distributed resonance patterns emerge in oscillatory networks once the dynamics of the oscillatory units become more than one-dimensional.
The network resonances are topology-specific and emerge at an intermediate frequency content of the input signals, between global yet homogeneous responses at low frequencies and localized responses at high frequencies. Our analysis reveals why these patterns arise and where in the network they are most prominent.
These results may thus provide general theoretical insights into how fluctuating signals induce response patterns in networked systems and simultaneously help to develop practical guiding principles for real-world network design and control.

\end{abstract}

\maketitle

\section*{Introduction}

Collective oscillatory network dynamics prevails in natural and technological systems alike, including neural and gene regulatory circuits, communication networks and AC power grids, among others \cite{strogatz2001, PikovskySynchronization, muresan2008, jahnke2014, karzbrun2014, filatrella2008, rohden2012,motter2013,motter2018,kirst2016,witthaut2016}.
A robust function is essential for all of these systems and relies on how their collective dynamics self-organizes and responds to external signals.
For example, modern power grids are driven dynamically by predictable or random signals, e.g. the fluctuating power feed-in from renewable energy sources, changes in consumer behavior on many time scales, power trading and failures of infrastructures, all of which cause fluctuations in grid frequency in a non-trivial, distributed way \cite{schaefer2018}.
The dynamics of all oscillatory systems essentially underlies system functions and all are externally driven, by planned interference signals, random fluctuations, unit or link failures etc. Yet it is not well known how oscillatory network systems respond to such perturbations.

To analyze, understand and predict the dynamics of coupled oscillators, networks of phase oscillators such as the paradigmatic Kuramoto model, are commonly employed across fields \cite{kuramoto1975,acebron2005,strogatz2000}.
Phase oscillator systems are preferred as generic models because they are simple to simulate, partly analytically accessible, and exhibit a wide range of collective dynamical phenomena observed in oscillatory systems, including synchronization and phase-locking, heteroclinic switching dynamics, collective irregular activity, and chimera states \cite{PikovskySynchronization,ashwin2005,neves2012,kotwal2017,bick2017}. 
Some knowledge about how such networks respond to external driving signals is already established. For instance, regularly and sparsely connected networks of phase oscillators exhibit two response regimes that depend on the frequency content of the driving signals \cite{zanette2004,zanette2005}. At low driving frequencies, the entire system dynamically responds homogeneously, i.e. all oscillators respond essentially identically. At high frequencies, the dynamical responses are localized on the network \cite{zanette2005} and response amplitudes decay with distance, similar to the findings for chains with single, constantly driven pacemaker units \cite{radicchi2006}. 
For the second-order power grid model, the network response to a local and static perturbation may be localized or delocalized depending on network topology and other parameters \cite{kettemann2016,wolff2018}.
However, an overarching theory of how networks respond systemically to dynamic, fluctuating inputs is missing to date such that the range of possible response patterns and the precise mechanisms underlying them are unknown in general and predicting them remains generally hard if not impossible.

In this article, we develop a theory of dynamic response patterns for oscillatory networks driven by fluctuating inputs. Evaluating a network-wide linear response theory as an explicit function of frequency content of the fluctuating signals, the network's interaction topology, and the exact location of driving and response nodes, we uncover complex distributed resonance patterns in oscillatory networks with two-dimensional unit dynamics, which are absent in standard phase oscillatory networks with one-dimensional units.  These network resonances emerge at intermediate frequency content of the input signals, between global yet homogeneous responses at low frequencies and localized responses at high frequencies.
Intriguingly, resonant responses may be an order of magnitude (greater than 10 times in our examples) larger than static responses that occur in the limit of zero frequencies in the identical network.
The theory furthermore explicates the origin and scaling of the localized as well as the homogeneous response patterns present, explaining and generalizing previous works \cite{zanette2004,zanette2005,kettemann2016,radicchi2006}. The finding and the explanation of the emergence of distributed resonance patterns together with the unifying theory we derive offers explanatory and predictive power across fluctuation-driven systems in nature and engineering.

\section*{Results}

\subsection{Model Class}

Consider a network of $N$ units $i\in \{1,\ldots,N\}$ with dynamics
\begin{equation}
\alpha_i\dfrac{d\theta_i}{dt}+\beta\dfrac{d^2\theta_i}{dt^2}=\Omega_i +\sum_{j}K_{ij}g(\theta_j-\theta_i) + F_i(t)
\label{eq:swing equation perturbed}
\end{equation}
of phases angles $\theta_i(t)$ and phase velocities  $d\theta_i(t)/dt$ at time $t$. Here $\Omega_i$ is the natural frequency (or acceleration) of unit $i$, and the units interact with coupling strengths $K_{ij}=K_{ji}\geq0$ via a coupling function $g(\cdot)$.
Core parameters $\alpha_i$ and $\beta_i$ in this model class set the relative influences of the rates of change of the phase angles and those of the phase velocities. The phenomena presented below for homogeneous  $\alpha_i\equiv \alpha $ and $\beta_i\equiv \beta$ stay qualitatively the same for inhomogeneous systems.
The model class \eqref{eq:swing equation perturbed} generalizes standard phase-oscillator networks \cite{acebron2005,strogatz2000} that are recovered for $\beta=0$. 
Taking generically both $\alpha,\beta>0$ and $g(\cdot)=\sin(\cdot)$ yields coupled second-order Kuramoto models characterizing AC power grids, the swing equation on networks \cite{filatrella2008, KundurPower, rohden2012}, on which we focus in the main part of this article.
For such systems, $\theta_i(t)$ describes the mechanical angle of a synchronous machine relative to a reference frame rotating at the nominal grid reference frequency of, e.g., $2\pi\times50$Hz in Europe including the United Kingdom (see \cite{filatrella2008, rohden2012} for details).
Normal stationary operation of a grid is given by a phase locked state, where (\ref{eq:swing equation perturbed}) exhibits a fixed vector $(\theta_1^{\ast},...,\theta_N^{\ast})$ of relative phases in the absence of any fluctuations ($F_i(t)\equiv 0$). The functions $F_i(t)$ represent the fluctuations in the production and consumption at unit $i$ and together constitute a high-dimensional, distributed dynamic driving signal to the network. How does the network collectively respond to such dynamic signals?

\subsection{Fluctuation-induced dynamic network responses}

\begin{figure*}[ht!]
\includegraphics[width=\textwidth]{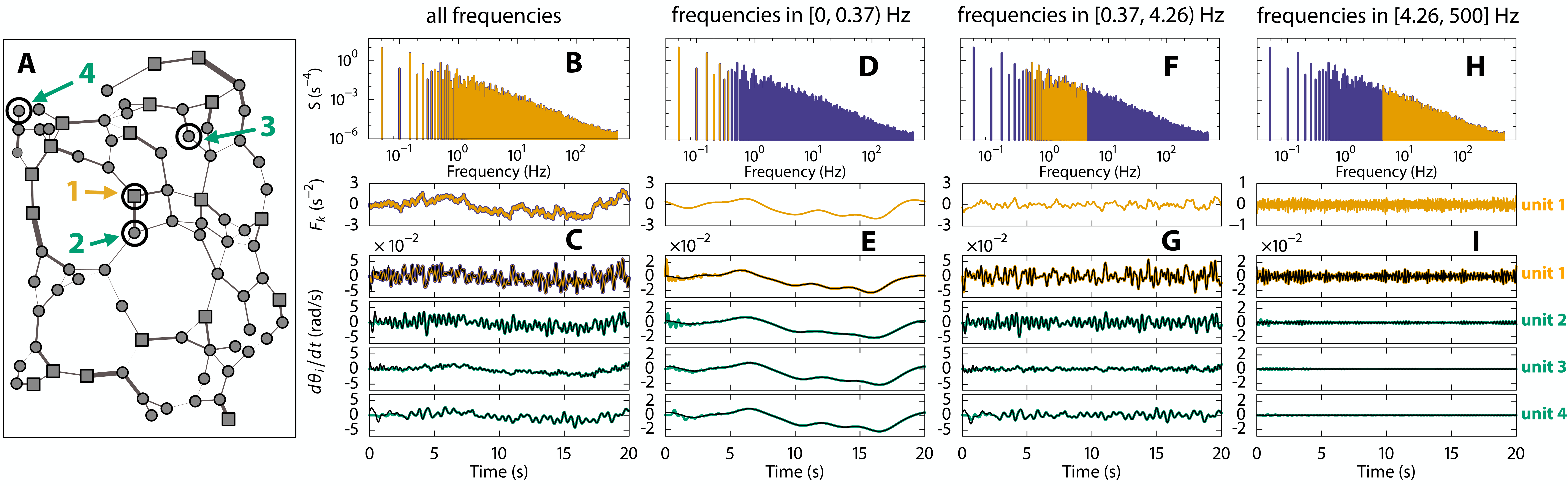}
\caption{
\textbf{Fluctuation-induced dynamic response patterns.}
A sample network (\textbf{A}) driven by Brownian noise (\textbf{B}) at a single unit $k=1$ exhibiting (\textbf{C}) complex dynamic response patterns that nonlinearly vary with frequency content of the input signal (\textbf{D}, \textbf{F}, \textbf{H}) as well as with distance between input and response unit (\textbf{E}, \textbf{G}, \textbf{I}). Three response classes emerge: homogeneous responses at low frequencies (\textbf{D}, \textbf{E}); at intermediate frequencies, spatiotemporally irregular patterns (\textbf{F}, \textbf{G}) that turn out to be characteristic for the network's interaction topology (see Eq.~\ref{eq:LRTsolution} below), and localized responses at high frequencies (\textbf{H}, \textbf{I}).
To identify frequency regimes, we selected specific frequency bands, yellow in (\textbf{D}, \textbf{F}, \textbf{H}), from the spectrum of the original noise realization (\textbf{B}), and all others displayed in purple.
Panels (\textbf{C}, \textbf{E}, \textbf{G}, \textbf{I}) display the time series of driving signal at unit $k=1$ (upper panels) and the response in the phase velocities (or frequencies) of the units $i\in \{1,\ldots,4\}$ (lower panels).
In (\textbf{C}, \textbf{E}, \textbf{G}, \textbf{I}) the time series of the reconstructed, band-selected signal, are displayed together with the responses (yellow for unit 1, green for units 2--4).
The analytic predictions, Eq.~\ref{eq:LRTsolution} (thin black lines) well predict the system responses obtained from direct numerical simulations. (For details of further settings, see Appendix.)}
\label{fig:network}
\end{figure*}

Observing the distributed network responses induced by a single local signal, $F_i(t)\sim\delta_{ik}$, already poses puzzles about how the responses depend on the driving and the response unit on the network topology, and on the frequency content of the signal, as illustrated in Fig.~\ref{fig:network}. Specifically, response to low-frequency inputs appear to be dynamic in time yet homogeneous across the network  (Fig.~\ref{fig:network}D and \ref{fig:network}E) and responses to high-frequency inputs seem to be localized (Fig.~\ref{fig:network}H and \ref{fig:network}I), whereas at intermediate frequencies, responses appear to be irregular in time and across units (Fig.~\ref{fig:network}F and \ref{fig:network}G), with severe consequences for the entire collective response pattern, compare Fig.~\ref{fig:network}B and \ref{fig:network}C.

To gain analytic insights and thereby intuition about such dynamic response
patterns, we start by focusing on the responses generated at one fixed (but arbitrary) unit $k$ by signals of fixed (arbitrary) frequencies $\omega$ and develop a linear response theory expanding in the amplitudes of the response signals into Fourier modes. In a second step, we consider more complex, distributed and non-periodic input signals below. Taking $F_i(t)= \varepsilon \delta_{ik} e^{\imath(\omega t+\varphi)}$ and expanding the phase responses to first order in the signal amplitude $\varepsilon$, we obtain
\begin{equation}
\theta_i^{(k)}(t)=\theta_i^{\ast}+\varepsilon\Theta_i^{(k)}(t)+\varepsilon\eta_i,
\end{equation}
where
\begin{equation}
\Theta^{(k)}_i(t)= A^{(k)}_i e^{\imath(\omega t+\Delta_i^{(k)})}
\label{eq:oscillatoryPhaseShift}
\end{equation}
denotes the oscillatory response at unit $i$ of amplitude $A^{(k)}_i$ and phase $\Delta_i^{(k)}$. $\eta_i=\eta\neq 0$ is an overall homogeneous phase shift that results from the dynamics of the average phase and intriguingly emerges despite zero average input.
For each input frequency $\omega$, a matrix equation
\begin{equation}
\left[-\omega^2\beta+\imath\omega\alpha+\mathcal{L}\right]\mathbf{\Theta}^{(k)}(t)
=\mathbf{F}^{(k)}(t),
\label{eq:response equation}
\end{equation}
thus describes the network response vector
$\mathbf{\Theta}^{(k)}(t)=\left(\Theta_1^{(k)}(t),\cdots,\Theta_N^{(k)}(t)\right)$ as a function of time.
The network response vector depends on the perturbation site $k$, the perturbation vector  $\mathbf{F}^{(k)}$  with only one non-zero component $F^{(k)}_i=\delta_{ik}\varepsilon e^{\imath(\omega t + \varphi)}$, parameters $\beta$ and $\alpha$,  and an weighted graph Laplacian matrix $\mathcal{L}$ with elements  $\mathcal{L}_{ij} := - K_{ij} \cos(\theta_j^* - \theta_i^*) $ for $i \neq j$ and $\mathcal{L}_{ii}=-\sum_{n=1; n\neq i}^N\mathcal{L}_{ni}$.
As $\mathcal{L}$ is real and symmetric, its $N$ eigenvalues are real, positive in normal operating states  where $\theta^*_j-\theta^*_i\leq\pi/2$ for all edges  $(i,j)$. Its eigenvalues are ordered as
 $\lambda^{[N-1]}> \cdots >\lambda^{[1]}>\lambda^{[0]}=0$, and its
eigenvectors $\{\mathbf{v}^{[0]},\cdots,\mathbf{v}^{[N-1]}\}$ form an orthogonal basis. The response vector
\begin{align}
\mathbf{\Theta}^{(k)}= e^{\imath(\omega t+\varphi)}\sum_{\ell=0}^{N-1}\dfrac{v^{[\ell]}_k}{-\beta\omega^2+\imath\alpha\omega+\lambda^{[\ell]}}\mathbf{v^{[\ell]}}.
\label{eq:LRTsolution}
\end{align}
thus provides a general first order prediction for the induced response patterns across all connected network topologies, compare ref. \cite{manik2014}.

Indeed, after a transient time set by the intrinsic dissipation time scale $1/\alpha$, this time-dependent vector function (\ref{eq:LRTsolution}), integrated across the signal's frequency content, well predicts the dynamic response patterns of the entire network, see, e.g., Fig.~\ref{fig:network}.
We emphasize that the prediction (\ref{eq:LRTsolution}) immediately generalizes to all network systems described by (\ref{eq:swing equation perturbed}) for arbitrary antisymmetric coupling function $g(\theta_j-\theta_i)$.
The asymmetry of $g(.)$ ensures a real and symmetric Laplacian matrix thus the orthogonality of Laplacian eigenvectors.

\subsection{Emergence of distributed resonance patterns}
The network response theory presented above provides an analytical expression of steady nodal responses (Eq.~\ref{eq:LRTsolution}), thus paving the way to a deeper understanding of the emergent complex response patterns across the network.
How does this response theory predict the different response regimes observed (Fig.~\ref{fig:network})? More specifically, how to explicitly extract the dependences of the network response patterns on the frequency of the driving signal, on the network interaction topology and on the graph-theoretic distance between driving and response units?
To reveal more detailed insights, we systematically analyze the amplitude $\omega A^{(k)}_i$ of the response in phase velocities $d\theta_i(t)/dt$ to a driving signal at unit $k$ across many orders of magnitude of the frequency and for all response units $i$ in a network. In this section we assume $\beta=1$ without loss of generality.

\begin{figure*}[ht!]
\centering
\includegraphics[width=0.8\textwidth]{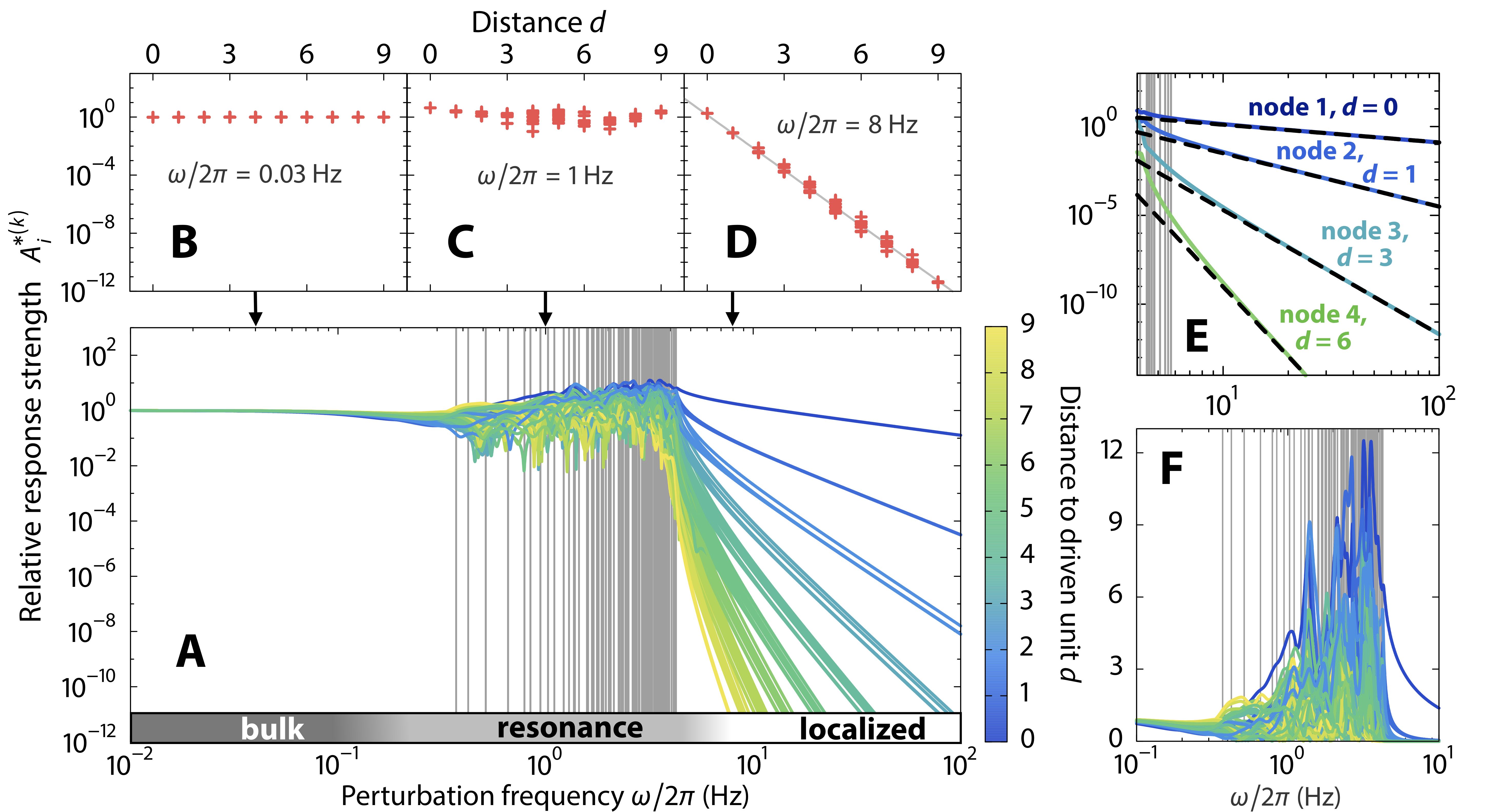}
\caption{
 \textbf{Emergence of resonances and prediction of distinct network responses.}
(\textbf{A}) The relative response strength $A^{\ast(k)}_i$ of each unit in the fluctuation-driven network illustrated in Fig.~\ref{fig:network}A vs. the signal frequency $\omega$, color-coded by the distance $d$ from the driven to the responding unit.
The eigenfrequencies are indicated by gray vertical lines and the three regimes of homogeneous bulk responses, heterogeneous resonant responses, and localized responses, by the gray-level gradient bar at the bottom.
(\textbf{B}), (\textbf{C}), and (\textbf{D}): The relative response strengths characteristically depend on distance, shown for individual frequencies representative for each of the three regimes.
(\textbf{E}) In the localized regime, the response amplitudes $A^{\ast(k)}_i$ for units 1 to 4 (marked in Fig.~\ref{fig:network}A) are well approximated by the analytic prediction (dashed lines), Eq.~\ref{eq:hi-freq approx}.
(\textbf{F}) Response strengths (plotted on linear scale) at resonance peaks may be an order of magnitude (here up to 12 times) larger than in the static response limit of $\omega \rightarrow 0$.
}
\label{fig:freq_dep}
\end{figure*}

Such responses in phase velocities reflect deviations of the units' frequencies of power grids from their nominal value $2\pi\times50$Hz and are thus more directly relevant than deviations in the phase angles themselves. Graphing the response strengths
\begin{align}
A^{\ast(k)}_i(\omega):=\dfrac{\omega A^{(k)}_i(\omega)}{\displaystyle\lim_{\omega'\rightarrow 0}\omega' A^{(k)}_i(\omega')}=N\alpha\omega A^{(k)}_i(\omega)\label{eq:DefRelativeResponseStrength}
\end{align}
relative to their low-frequency limit clearly indicates the three frequency regimes of characteristic responses, see Fig.~\ref{fig:freq_dep}A.

Dynamic resonance patterns emerge if the signal frequencies $\omega$ are of the same order of magnitude as the eigenfrequencies 
\begin{equation}
\omega^{\ast}_n=\sqrt{\lambda^{[n]}-\dfrac{\alpha^2}{4}},\quad \text{for } \ n \in \{1,\cdots,N-1\}.
\end{equation}
The system responds with a distributed, network- and frequency-specific pattern given by Eq.~\ref{eq:LRTsolution} (Fig.~\ref{fig:freq_dep}A, \ref{fig:freq_dep}C and \ref{fig:freq_dep}F). Network resonance patterns are typically heterogeneous, exhibiting no monotonicity in frequency or with respect to the topological distance between the signal and the response unit.
In contrast to resonance phenomena in low-dimensional oscillatory systems common in physics and nonlinear dynamics, the response patterns are node-specific in a highly non-trivial way. Depending on an \emph{overlap factor} $v^{[\ell]}_k v^{[\ell]}_i$ of the responding node $i$ with the driven node $k$ in eigenmode $\ell$, the contribution from the various eigenmodes can be positive, negative or zero, which leads to different canceling or accumulating total responses for each frequency $\omega$. The overlap factor acts as a node-specific weight in the eigenvector superposition, thus intuitively explaining the complex distributed dynamic response patterns. Intriguingly, the response amplitude at resonance peaks may be an order of magnitude larger than for static responses appearing in the limit of $\omega \rightarrow 0$, cf. Fig.~\ref{fig:freq_dep}F.

For frequencies substantially below the eigenfrequencies $\omega^{\ast}_n$ all units exhibit similar response strengths, in particular independent of the topological distance between signal input and response sites (Fig.~\ref{fig:freq_dep}A and \ref{fig:freq_dep}B). At the low frequency limit $\omega\rightarrow0$, $A_i^{*(k)}$ approaches $N\alpha$, a constant containing only intrinsic network parameters, not encoding any topology dependence. This explains the observation of homogeneous responses in Fig.~\ref{fig:network}D and \ref{fig:network}E.

For sufficiently large driving frequencies $\omega \gg \omega_n^{\ast}$ the response strengths decay as a power law with frequency. Intriguingly, this decay is essentially exponential over the topological distance between signal input and unit response sites (Fig.~\ref{fig:freq_dep}A, \ref{fig:freq_dep}D).
Writing the response amplitudes as rational functions of $\omega$, with $\Phi_{ki}^{[2N-1-j]}$ as coefficients of $\omega^{2j}$ in the numerator of $\mathrm{Re}\left(\Theta^{(k)}_i(t)\right)$, we obtain the first non-zero term with the highest power of $\omega$ in the numerator to be $\Phi_{ki}^{[d]}\omega^{4N-2-2d}$, where $d:=d(k,i)$ is the topological distance between the driving unit $k$ and responding one $i$. The highest power $\omega^{4N}$ appearing in the denominator thus yields the exact asymptotic expansion
\begin{align}
 A^{\ast(k)}_i(\omega)\sim N\alpha\left|\Phi_{ki}^{[d]}\right|\omega^{-2d-1}.
\label{eq:hi-freq approx}
\end{align}
as $\omega \rightarrow \infty$ via definition (\ref{eq:DefRelativeResponseStrength}). 
Here the frequency dependence is encoded solely in the term $\omega^{-2d-1}$. The topological distance $d$ appears explicitly twice and $\Phi_{ki}^{[d]}$ encodes overall topological properties of the network as well as the identity of the driving and the responding nodes.
This result \eqref{eq:hi-freq approx} is based on exact zeros of lower-order coefficients that appear because the relevant matrix element of the $m$-\emph{th} power of the Laplacian is zero, $\left(\mathcal{L}^m\right)_{ki}=0$, if $m<d(k,i)$ as there exists no path from $k$ to $i$ of length smaller than $d(k,i)$. Consequently, for any given frequency in this regime, the response strengths decay approximately exponentially with the distance.

\begin{figure}[ht!]
\centering
\includegraphics[width=\columnwidth]{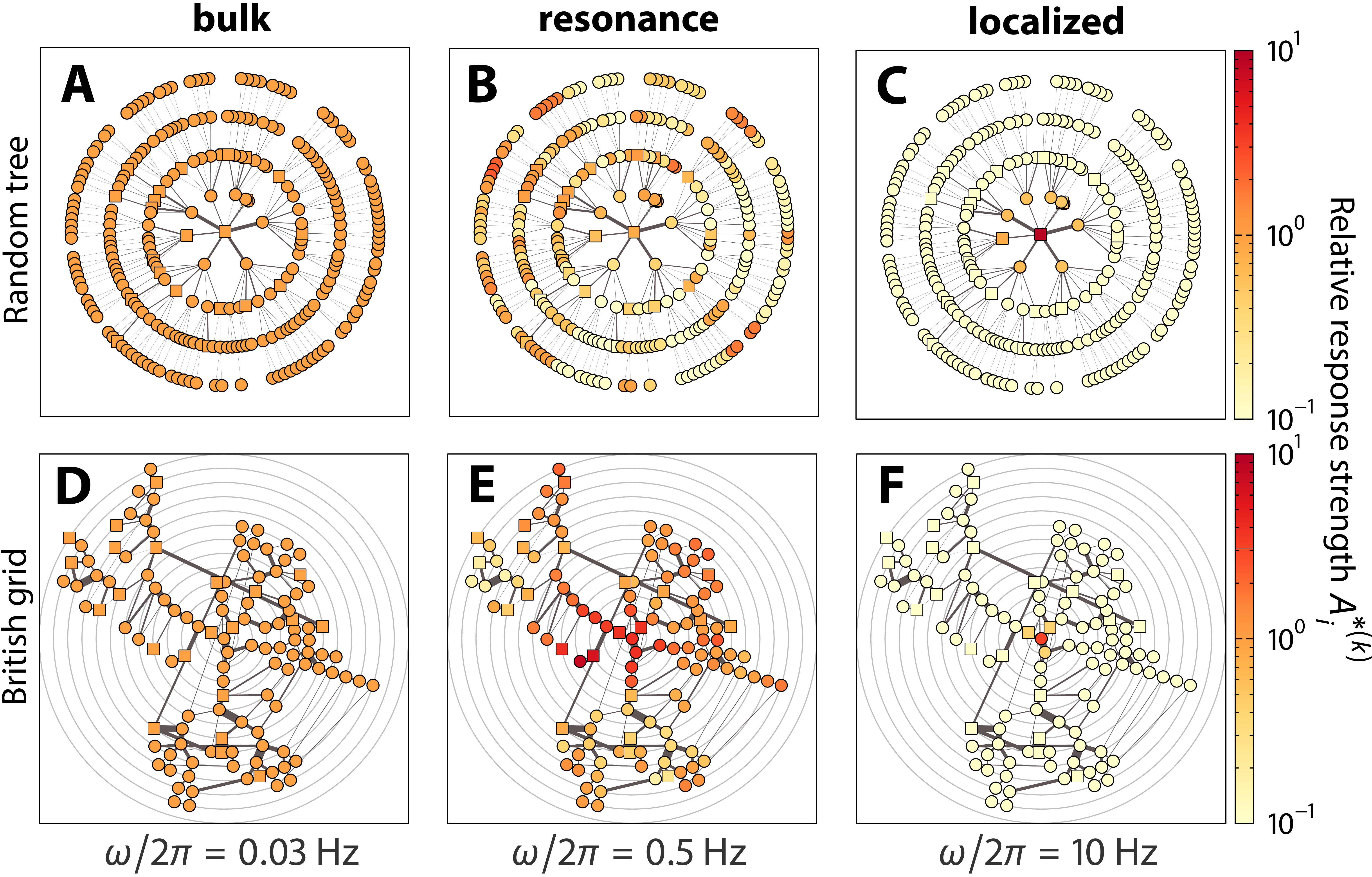}
\caption{
\textbf{Generality of response patterns.}
For two exemplary networks, the response patterns for three signals of frequencies representing the three response regimes are indicated by their relative response strengths $A^{\ast(k)}_i$ (color coded). Response patterns for (\textbf{A}-\textbf{C}), a random tree with $N=264$ and (\textbf{D}-\textbf{F}), the topology of the British high-voltage transmission grid with $N=120$, both reordered according to graph-theoretic distance from site of driving unit. For both networks, the driven unit is placed at the center with all units displayed on circles with their radii proportional to topological distance (gray concentric rings) in (\textbf{D}-\textbf{F}).}
\label{fig:patterns}
\end{figure}

We emphasize that the above analysis about the three regimes of response patterns holds without any assumptions on the network topology, thus is general for arbitrary interaction structures. See Fig.~\ref{fig:patterns} for an illustration of the generality of the response regimes across different networks.

\subsection{Fluctuation-induced distributed resonances}
So far, we have systematically studied the patterns of network responses to trigonometric, sufficiently weak fluctuating signals affecting just one unit in the network.
What can we say about the dynamic response patterns in networks with stronger, more irregular and more distributed driving signals?

For input signals that are both distributed across the network and contain a broad range of frequency components ($\omega>0$), linear response theory predicts the units' responses as a linear superposition across input locations and frequencies via Eq.~\ref{eq:LRTsolution}.
Specifically, the phase velocity at unit $i$ reads 
\begin{align}
\text{\footnotesize$
\dot\theta_i^{(\bm{\kappa})}=\displaystyle\sum_{k\in\kappa}\Bigg(\dfrac{\varepsilon_{0,k}}{N\alpha}+\sum_{n}\imath\varepsilon_{n,k}\omega_{n,k}\sum_{\ell=0}^{N-1}\dfrac{v^{[\ell]}_kv^{[\ell]}_ie^{\imath(\omega_{n,k} t+\varphi_{n,k})}}{-\beta\omega_{n,k}^2+\imath\alpha\omega_{n,k}+\lambda^{[\ell]}}\Bigg),$}
\label{eq:response noisy}
\end{align}

where $\bm{\kappa}$ denotes the set of all units driven by fluctuating signals and $n,k$ labels the dominant  Fourier modes with non-zero frequencies at each such unit $k$. The constant drift speed $\frac{\varepsilon_{0,k}}{N\alpha}$ results from the zero-frequency component with magnitude $\varepsilon_{0,k}$ in the signal at node $k$.

Generically, our theory, Eq.~\eqref{eq:response noisy}, has strong predictive power for all combinations of differently fluctuating inputs distributed across the network (see Fig.~\ref{fig:noisy}A-\ref{fig:noisy}D and more details in Appendix). 
Moreover, the eigenvector-based overlap factor $v^{[\ell]}_kv^{[\ell]}_i$ in Eq.~\eqref{eq:response noisy} clearly demonstrates how frequency components in the driving signal that are close to the eigenfrequencies induce network-wide resonances in a specific, highly topology-dependent way.

\begin{figure}[h!]
\centering
\includegraphics[width=0.5\textwidth]{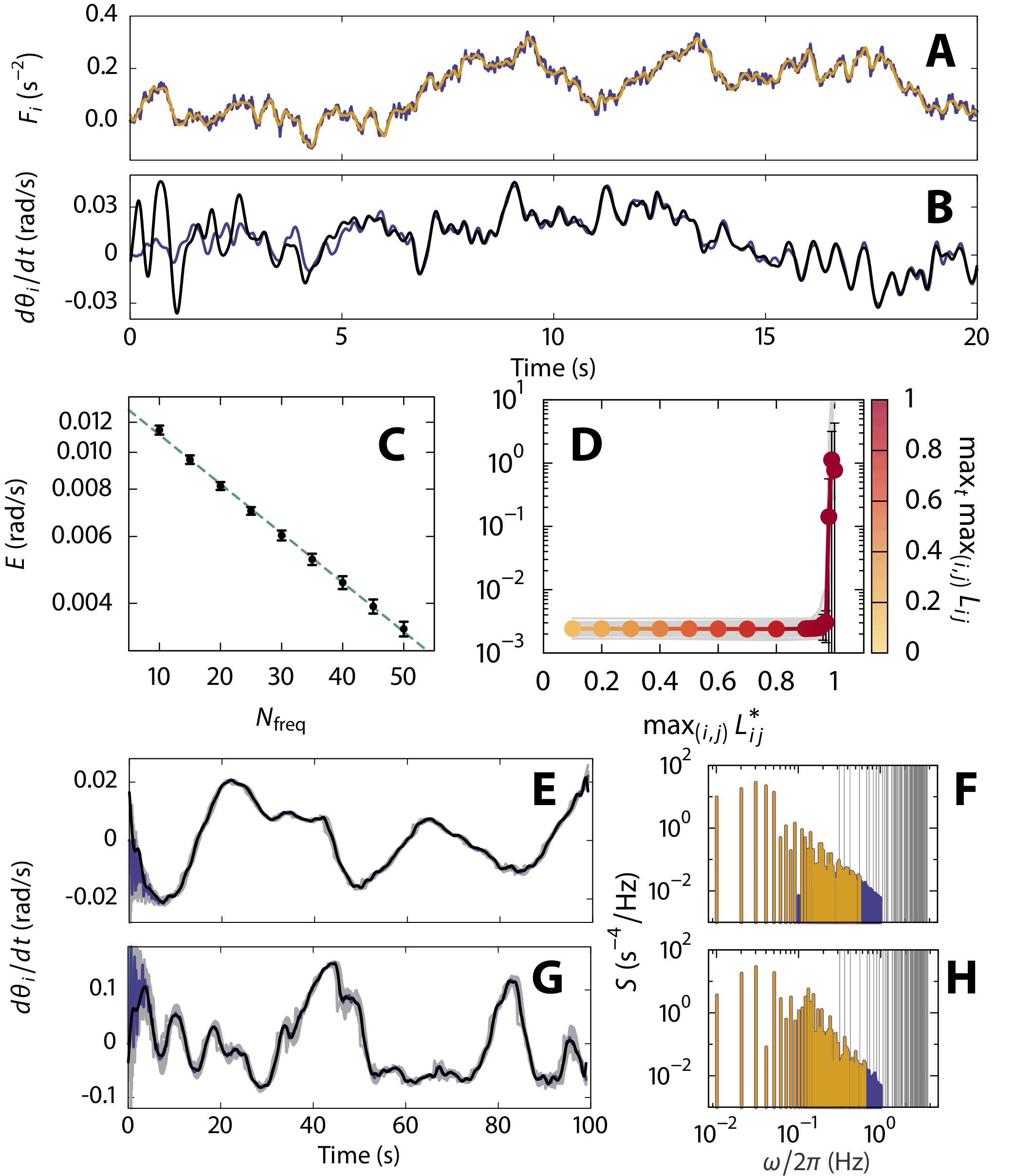}
\caption{
\textbf{Predicting response patterns of fluctuation-driven networks.}
(\textbf{A}) Time series (purple) of a noisy signal at one unit and (\textbf{B}) frequency response at another unit in the network of Fig.~\ref{fig:network}A, with every unit driven by independent Brownian noise. Response prediction (black) is based on $50$ dominant Fourier modes of the input signal at each unit, which, after a transient phase, is very close to the numerical response (purple) to the original signals. The signal reconstructed by the $50$ dominant Fourier modes is displayed in yellow in (\textbf{A}).  (\textbf{C}) Prediction error (definition see Appendix) decreases exponentially with the number of Fourier modes used for prediction (errorbars indicate standard deviation of prediction error across $70$ realizations of noisy signals). (\textbf{D}) Prediction error as a function of line load $L_{ij}:=|\sin(\theta_j-\theta_i)|$. Linear response theory well predicts responses until the maximum load of all transmission lines exceeds 95\% load of transmission lines; breakdown of linear response theory indicated by drastic increase of error if the system is almost fully loaded at the operating point.
(\textbf{E}, \textbf{G}) Example response time series of phase velocity (i.e. grid frequency) in the network of Fig.~\ref{fig:network}A for (\textbf{E}) a fluctuating wind power input signal, and (\textbf{G}) a fluctuating photovoltaic panel power recording with respective power spectral density $S(\omega)$ (\textbf{F}) and (\textbf{H}). The gray vertical lines indicate the system's eigenfrequencies. The response time series of one unit is highlighted in purple (analytic prediction as above, marked in black) and of all others in the network in gray. The prediction is made based on $50$ dominant Fourier modes of the input signal, highlighted in yellow in (\textbf{F}) and (\textbf{H}).}
\label{fig:noisy}
\end{figure}

To illustrate how distributed network resonances are induced by a noisy fluctuation signal due to its resonant frequency components, we computed the response time series for a power grid model network with the topology employed for Fig.~\ref{fig:network}A. The fluctuating signals were experimentally recorded from renewable energy sources. The results indicate similarities as well as notable differences between the responses to the recordings from photovoltaic and from wind power sources. Specifically, we find that for both time series samples, fluctuations in wind power and fluctuations in the power from a photovoltaic panel (see Appendix), predictions are highly accurate after a transient time (Fig.~\ref{fig:noisy}E, \ref{fig:noisy}G). In both sample time series of power fluctuation, the low-frequency components are dominant (Fig.~\ref{fig:noisy}F, \ref{fig:noisy}H) and induce low-frequency fluctuations that are homogeneous across the network (Fig.~\ref{fig:noisy}E, \ref{fig:noisy}G). In addition, the stronger and faster input fluctuations in photovoltaic power generate responses that are highly precise in time yet more inhomogeneous across the network (Fig.~\ref{fig:noisy}G, \ref{fig:noisy}H). This phenomenon originates from the resonances induced by the fluctuation signal: the Fourier spectrum of the photovoltaic power fluctuation contains relatively stronger frequencies close to the resonance frequencies than the Fourier spectrum of the wind power fluctuation (see Figs.~\ref{fig:noisy}E-\ref{fig:noisy}H).

\section*{Discussion}

To conclude, we established a theory of \textit{dynamic} response patterns in fluctuation-driven oscillatory network systems by systematically analyzing the response patterns in every dynamical range of signal's frequency content. 
We revealed the emergence of the distributed resonance patterns in the intermediate frequency regime and demonstrated that they result from the second dynamic variable of individual oscillators, which are known to be absent in Kuramoto oscillator systems \cite{zanette2004,zanette2005}.
Resonance patterns are topology-dependent, heterogeneous, nonlinear and non-monotonic in distance. 
For high and low frequency regimes, we analytically derived the generic response patterns, i.e. the homogeneous patterns at low frequencies and the localized patterns at high frequencies, by rigorously extracting the response' dependence on distance, providing solid theoretical support and broad generalization for previous works \cite{zanette2004,zanette2005,kettemann2016,radicchi2006}.

The reported findings on network response patterns exhibit a number of implications for the operation and control of real-world network systems in biology and engineering. For the example of power grids, it is often believed that the AC frequency is homogeneous across the entire interconnected power system, thus the notion ``grid frequency'' in power grid literature e.g. \cite{short2007}. In strong contrast, our results demonstrate that the frequency responses are indeed highly heterogeneous, therefore, ``grid frequency'' makes sense only in the regime of slow-changing driving signals. 
In particular, anomalous broad-range fluctuations of grid frequencies \cite{schaefer2018} may in part result from the combination of different response patterns to various input fluctuations across grid nodes.

The phenomenon of resonance was known before from complex dynamic systems if they were driven by external signals that act globally on all units. For instance, in 1981, Benzi reported that stochastic resonance may arise from global, homogeneous periodic forcing of stochastic dynamical systems \cite{benzi1981,wiesenfeld1995,benzi2009}. Even earlier (non-stochastic) resonance of molecular excitations was triggered by incident photons with a frequency matching the energy of an electronic transition of the molecule and was exploited in resonance Raman spectroscopy \cite{spiro1977}. Already in the 1950s, related fundamental phenomena were exploited to reveal energy spectra, e.g. of atomic nuclei \cite{wigner1956}. The work presented here demonstrates network-wide distributed resonances that emerge in oscillatory networks due distributed fluctuations impinging distinctly on the different individual units. Previously found zero-frequency `resonances' in networks of (Kuramoto) phase oscillator networks exhibit a single response peak consistently appearing at zero frequency, independent of the network topology and the system's eigenfrequencies \cite{zanette2004,zanette2005}. In stark contrast, the resonances revealed and explained above, are strongly specific to the interaction topology.

Taken together, our results indicate that the second dynamical dimension of an oscillator (its variable local frequency) present in the general model class (Eq.~\ref{eq:swing equation perturbed}) induce characteristic dynamic and distributed resonance patterns strongly depending on which units are driven by fluctuations and on how the entire network is connected. 
The topology- and the explicit distance-dependence as well as the frequency-dependence of the responses revealed here may not only provide a general theoretical prediction of how networks respond to distributed input fluctuations. It may also find a range of useful applications. For instance, given typical frequency content of fluctuation signals at specific units, our theory enables the identification of the most susceptible units in the network, and thereby suggest dynamically motivated design constraints and the prevention or mitigation of systemic risks of functional failure.

\begin{acknowledgements}
We thank Malte Schr\"oder and Debsankha Manik for insightful discussions. 
We gratefully acknowledge support from the Federal Ministry of Education and Research (BMBF grant no. 03SF0472A-F), the Helmholtz Association (via the
joint initiative “Energy System 2050—a Contribution of the Research Field Energy” and grant no. VH-NG-1025), and the German Science Foundation (DFG) by a grant toward the Cluster of Excellence “Center for Advancing Electronics Dresden” (cfaed).
\end{acknowledgements}

\section*{APPENDIX: Settings for figures}

\paragraph*{Settings for Fig.~\ref{fig:network}:}
Random geometric topology in (\textbf{A}) generated according to a growth model of power grids \cite{schultz2014}, where the cost-vs-redundancy trade-off parameter $r = 0$, meaning line redundancy is disregarded in the network growth. Squares represent generators with rated power inputs $\Omega=30s^{-2}$ and disks represent consumers with power consumption $\Omega=-10s^{-2}$. Number of units $N=80$. For all transmission lines $K=100s^{-2}$, and for all units $\alpha=1s^{-1}$.
The fluctuation signal is generated by Wiener process with drift $0$ and volatility $1$. 

\paragraph*{Settings for Fig.~\ref{fig:noisy}:}
In (\textbf{C}, \textbf{D}), to assess the quality of the prediction for phase velocity, we define the prediction error as
\begin{align}
E=\left\langle\left(\dot\theta_i^{\text{LRT}}(t)-\dot\theta_i^{\text{num}}(t)\right)^2\right\rangle_{t,i}^{\frac{1}{2}},
\label{eq:error}
\end{align}
where $\dot\theta_i^{\text{LRT}}$ and $\dot\theta_i^{\text{num}}$ are the phase velocity responses obtained from the above linear response theory (\ref{eq:response noisy}) and direct numerical simulations, respectively. The angular brackets indicate the average over time and over all units in the network.
(\textbf{D}) By proportionally amplifying the power generation and consumption $\Omega_i$ with an incrementally increasing factor $r$: $\Omega_i\rightarrow r\Omega_i$, we systematically increase the line loads $L_{ij}^*=|\sin(\theta^*_j-\theta^*_i)|$ for every link $(i,j)$ at the steady state and test the prediction error for the same noise time series as in Figs.~\ref{fig:noisy}A-\ref{fig:noisy}B. In the prediction $50$ dominant Fourier components are included. The data points are color-coded by the maximum line load in the network under perturbation in a time interval $t=20s$. The errorbars stand for the standard deviation of prediction error over the network. The time average of the prediction error for individual nodes are plotted as light grey lines.
(\textbf{E}-\textbf{H}) The recordings of the fluctuating power output from renewable energy sources, wind turbines and PV panels are obtained from and available under Ref.~20 of \cite{anvari2016}. In the computation of network responses we use a coarse-grained model of power grids, with each node represents a coherent sub-grid.

\bibliography{Resonance}

\end{document}